\documentclass[journal=jpcl,manuscript=article]{achemso}

\usepackage{lineno}
\usepackage{amssymb}
\usepackage{xcolor}
\usepackage{siunitx}
\usepackage[hidelinks]{hyperref}

\usepackage[version=3]{mhchem} 

\author{Timothy S. Groves}
\email{timothy.groves@chem.ox.ac.uk}
\affiliation{Physical and Theoretical Chemistry Laboratory, University of Oxford, Oxford, UK}
\author{Susan Perkin}
\email{susan.perkin@chem.ox.ac.uk}
\affiliation{Physical and Theoretical Chemistry Laboratory, University of Oxford, Oxford, UK}

\title[An \textsf{achemso} demo]
  {Wave mechanics in an ionic liquid mixture}

\begin{document}

\begin{abstract}

Experimental measurements of interactions in ionic liquids and concentrated electrolytes over the past decade or so have revealed simultaneous monotonic and oscillatory decay modes. These observations have been hard to interpret using classical theories, which typically allow for just one electrostatic decay mode in electrolytes. Meanwhile, substantial progress in the theoretical description of dielectric response and ion correlations in electrolytes has illuminated the deep connection between density and charge correlations and the multiplicity of decay modes characterising a liquid electrolyte.  The challenge in front of us is to build connections between the theoretical expressions for pair correlation functions and the directly measured free energy of interaction between macroscopic surfaces in experiments. Towards this aim, we here present measurements and analysis of the interactions between macroscopic bodies across a fluid mixture of two ionic liquids of widely diverging ionic size. The measured oscillatory interaction forces in the liquid mixtures are significantly more complex than for either of the pure ionic liquids, but can be fitted to a superposition of two oscillatory and one monotonic mode with parameters matching those of the pure liquids. We discuss this empirical finding, which hints at a kind of wave mechanics for interactions in liquid matter.
\end{abstract}

\newpage
\section{Introduction}
The nano-scale structure and its relation to particle interactions in ionic liquid mixtures is important for many applications ranging from directed synthesis to the optimisation of electrolytes in batteries\cite{Welton,Mao:2019aa,Fayer2020}. In this article we present new experimental measurements of the interactions between macroscopic objects immersed in a mixture of two ionic liquids, measured using a surface force balance (SFB). Our results for mixtures of ionic liquids are quite complex compared to those for pure ionic liquids, so we begin with a brief discussion of the fundamental relation between measured interaction force between macroscopic surfaces or particles and the bulk liquid's structure and interactions. Theoretical relationships of this sort are already well established in general terms, but become complicated quickly with increasing complexity of the fluid and so are not always easily applied to realistic experimental scenarios.

\begin{figure}[h!]
	\centering
	\includegraphics[scale=1]{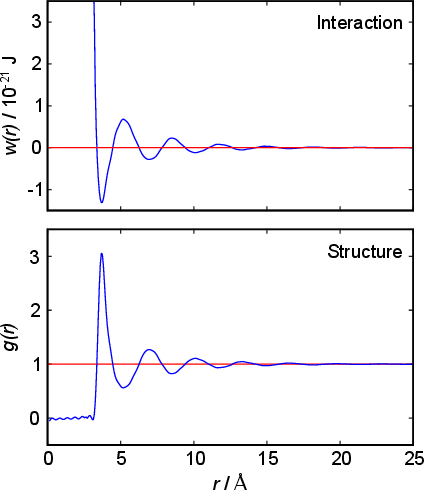}
	\caption{Bottom: Radial distribution function, $g(r)$, for liquid argon at 85~K, as measured by neutron scattering measurements by Yarnell \textit{et al.}\cite{yarnell1973_argon}{}. Top: pair potential of mean force, $w(r)$, derived from the measured $g(r)$ according to equation~\ref{eq:ppmf}. Both functions reveal a single damped oscillatory decay mode  with a wavelength close to the argon atomic diameter.}
	\label{fig:argon_g_w}
\end{figure}

The intimate relationship between liquid structure and interactions can be illustrated by reference to the simple case of liquid argon, for which direct measurements of the pair distribution function, $g(r)$, have been obtained from neutron scattering measurements\cite{yarnell1973_argon}{} as reproduced in figure~\ref{fig:argon_g_w}. In general, $g(r)$ is related to the pair potential of mean force, $w(r)$, in a 1-component atomic fluid by:
\begin{equation}
	g(r) = e^{-\beta w(r)}
	\label{eq:ppmf}
\end{equation}
where $r$ is the distance between two particles\cite{KjellanderBook}. We see that $g(r)$ and $w(r)$ are comprised of a decaying oscillation, which reflects intermolecular repulsions and packing constraints as expected for a simple fluid at high density\cite{Widom1967}. At low density, by contrast, excluded volume and many-body effects become insignificant and $w(r)$ tends towards $u(r)$, the mean field pair potential; oscillations of the sort apparent in figure~\ref{fig:argon_g_w} then disappear and intermolecular attractions dominate the behaviour\cite{KjellanderBook}.  

Detailed calculations for simple atomic fluids capture this experimental behaviour well\cite{Hansen_book,evans1993_liquidstate}. With the only length scales arising from  particle size and density, it is found that $g(r)$ will contain one term, or \textit{mode}. This mode may be oscillatory at high densities, dominated by \textit{density correlations}, or monotonic at low densities. The transition from monotonic to oscillatory decay in $g(r)$ extends beyond the critical point and the cross-over has been called the Fisher-Widom line\cite{fisher1969_widom}{}.

Most real liquids are far more complex than liquid argon; they cannot be parameterised simply by density and radius alone. Instead one would need to consider parameters capturing geometry, charge, charge distribution, polarizability, etc. in order to build a full picture of the structure and interactions within the fluid. Most particularly, electrostatic interactions arising from Coulomb interactions between ionic species in the liquid act over long range and so give rise to significant deviations from mean-field behaviour even at very low concentration. The most famous attempt to quantify the charge distribution in electrolytes is the Debye-H\"uckel theory which is built from the Poisson-Boltzmann equation, both of which involve strong approximations\cite{Onsager1933}. The key result of the Debye-H\"uckel theory is that the charge distribution and electrostatic potential away from a charged particle decay in a plain exponential manner (compared to the Coulomb power-law decay without `screening'). Soon after the Debye-H\"uckel theory, statistical analysis by Kirkwood of electrolyte properties over a wide range of concentrations predicted the fascinating result that that the charge-charge distribution function switches from plain exponential decay to a damped oscillatory decay at high charge density.  Although reminiscent of the crossover in particle density decay at the Fischer-Widom line mentioned above, these electrostatic charge-waves have a different origin and the boundary between plain exponential and oscillatory charge decay has been called the Kirkwood line.\cite{Kirkwood1936,Evans1994_ions,kjellander2019} 

Although the concepts of Debye screening (at low ion concentration) and the Kirkwood crossover from monotonic to oscillatory charge-charge correlations at high ion concentration are useful heuristic ideas for interpreting interactions in electrolytes, they cannot fully capture real electrolyte behaviour because they arise from theories which do not take into account the correlation between all ions in the electrolyte.  A theoretical framework addressing this issue is the \textit{dressed ion theory}\cite{Kjellander1994}, which has resulted in many important insights. A key result in the analysis of interactions in electrolytes relevant to our discussion here is that \textit{multiple decay modes} contribute additively to $g_{ij}(r)$, $w_{ij}(r)$, to the pair correlation function, $h_{ij}(r) = g_{ij}(r)-1$, and to the screened Coulomb potential, $\phi_{i}(r)$\cite{kjellander2019}. The \textit{decay modes} for the pair correlation function take the form:
\begin{multline}
	h_{ij}(r) \sim -\frac{1}{4\pi\epsilon_0k_BT} \left[\frac{q_i^{eff}(\kappa_1)q_j^{eff}(\kappa_1)}{\mathcal{E}_r^{eff}(\kappa_1)}\frac{e^{-\kappa_1 r}}{r} + \cdot \cdot \cdot \right] \\-  \frac{1}{2\pi\epsilon_0k_BT} \left[\frac{q_i^{eff}(\kappa_R)q_j^{eff}(\kappa_R)}{\mathcal{E}_r^{eff}(\kappa_R)}\frac{e^{-\kappa_R r}}{r}\cos(\kappa_I r + \phi)  + \cdot \cdot \cdot \right]
	\label{eq:hij}
\end{multline}
where the first bracketed term on the right implies that there are other monotonic Yukawa (\textit{i.e.} $\frac{1}{r}e^{-r}$) terms, and the second bracketed term on the right implies further oscillatory Yukawa terms\cite{kjellander2019}. $q_i^{eff}(\kappa)$ is a mode-specific ($\kappa$ dependent) effective charge for particle $i$, $\mathcal{E}_r^{eff}(\kappa)$ is the effective relative dielectric permittivity (also $\kappa$ dependent) and $\phi$ is a phase shift for the oscillation. Equation~\ref{eq:hij} applies to spherical ions, but similar general expressions describe the decay for any geometry. (The same terms appear in $w_{ij}(r)$, which can be seen by noting that asymptotically $w_{ij}(r) = -k_BTh_{ij}(r)$ from equation~\ref{eq:ppmf}.)  It is seen that each monotonic term is described by a single decay parameter (the first one is $\kappa_1$ above), and each oscillatory term is described by two parameters (e.g. $\kappa_R$ and $\kappa_I$ above). These decay parameters are solutions to the general equation for $\kappa$\cite{kjellander2016,kjellander2019}:
\begin{equation}
	\kappa^2 =  \frac{\sum_in_i^bq_iq_i^{\ast}}{\mathcal{E}_r^{\ast}(\kappa)\epsilon_0}
	\label{eq:kappa}
\end{equation}
where $n_i^b$ is the density of ion $i$ in the bulk electrolyte, $q_i$ is its bare charge, $q_i^{\ast}$ is its renormalised charge, and $\mathcal{E}_r^{\ast}(\kappa)$ is the dielectric factor\cite{kjellander2019}. Equation~\ref{eq:kappa} is an equation for $\kappa$ with many solutions, leading to the many decay modes in equation~\ref{eq:hij}. The solutions can be real, in which case the mode in the correlation function will be a monotonic Yukawa function with decay length $1/\kappa$, or pairs of complex conjugates giving rise to an oscillatory Yukawa term with decay length $1/\kappa_R$ and wavelength $2\pi/\kappa_I$. Equation~\ref{eq:kappa} reduces to the expression for the Debye-H\"uckel screening parameter, $\kappa_{DH}^2 =  \frac{\sum_in_i^bq_i^2}{\epsilon_r\epsilon_0k_BT}$, as $n_i^b \rightarrow 0$ and only in this limit is there just one decay mode (characterised by the Debye-H\"uckel  screening lenghth, $ \lambda_{DH} = 1/\kappa_{DH}$). \textit{In all real electrolytes, there is more than one decay mode and more than one screening length.}

The decay modes contributing to the pair correlation function also feature in the potential of mean force (per unit area) between two \textbf{macroscopic surfaces}, $W_{I,II}(D)$, as relevant to our experiments with a surface force balance (SFB)\cite{kjellander2016,Kjellander2018}. That is to say:
\begin{equation}
	W_{I,II}(D) \sim \left[Ce^{-\kappa_1D} + \cdot \cdot \cdot \right] + \left[C'{e^{-\kappa_R D}}\cos(\kappa_I D + \phi)  + \cdot \cdot \cdot \right]
	\label{eq:WI2D}
\end{equation}
where, again, the two brackets indicate that more terms of the monotonic and oscillatory decaying type exist. Each term, or mode, is characterised by a decay parameter (and, for the oscillatory mode, a wavelength) which are identical to the parameters which define the bulk pair correlations and pair potential of mean force, as above. The constants $C$, $C'$ etc depend on the effective charge on the surfaces. \\
Since the terms in $W_{I,II}(D)$ decay at different rates, the overall value of $W_{I,II}(D)$ may be dominated by different modes at different ranges of $D$. That is to say, as two planar surfaces approach one another from large distances, they may go through regions of $D$ where the interaction is dominated by a monotonically increasing interaction and other regions where the interaction is oscillatory. Indeed, many past experiments with any kind of electrolyte, ranging from dilute electrolytes to pure ionic liquids can be interpreted in this framework. 

\begin{figure}[h!]
	\centering
	\includegraphics[scale=1]{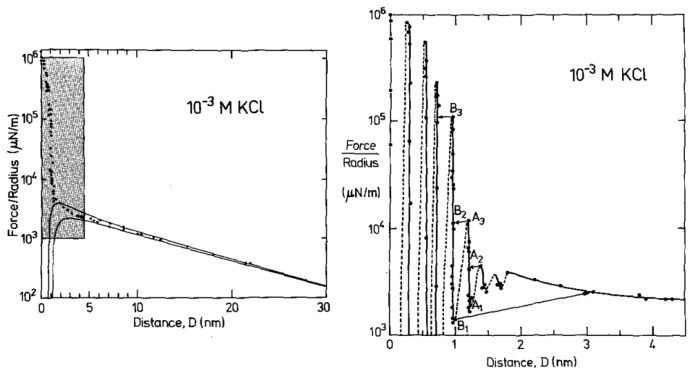}
	\caption{Interaction force between two macroscopic crossed cylinders across a thin film of $10^{-3}$~M~\ce{KCl}.
	Force data is normalised by cylinder radius, $F/R$, which is proportional to $W_{I,II}(D)$.
	The left hand panel shows the full interaction over 30~nm of surface separation. 
	The region highlighted in gray is shown in the right hand panel.
	The data shows two clear modes, a long range monotonic mode and a short range oscillatory mode.}
	\label{fig:israelachvili_data}
\end{figure}

We illustrate this interpretation of $W_{I,II}(D)$ by inspection of three example measurements. First, let us consider the measured interaction force across dilute aqueous electrolyte solutions; one of the first examples of such a measurement, for   $10^{-3}$~M~\ce{KCl} aqueous solution by Pashley and Israelachvili, is reproduced in figure~\ref{fig:israelachvili_data}.  Similar measurements have been made more recently, revealing similar key features\cite{hallett2023}. At large surface separations a monotonic repulsive interaction acts between the two charged surfaces; at small surface separations the interactions are dominated by an oscillating interaction with a wavelength of $\approx 0.25$~nm. Within the framework above, two \textit{modes} are apparent. The monotonic mode originates mostly from the ion correlations and is similar to the Debye-H\"uckel prediction at low ionic concentration but not precisely the same, and the oscillatory mode is due primarily to density correlations between the solvent molecules. The reason for the words `mostly' and `primarily' in the previous sentence is that, in fact, \textit{all correlations in the electrolyte contribute to each of the decay modes}. That is to say, the decay modes cannot -- in general -- be separated into terms which arise purely from charge-charge correlations, density-density correlations, etc.\cite{Kjellander2018}.  However, in some cases (such as a very dilute electrolyte) it is possible to provide a physical interpretation of the `origin' of each mode in terms of the components of the electrolyte. The contrast between this description and the usual terminology of electrostatic double-layer forces and hydration forces may seem a purely semantic distinction for this example, but as we shall see later the interpretation of interactions in more complex mixtures or higher electrolyte concentration cannot be accommodated without the more precise formulation.  

A second example measurement to interpret in light of the dressed ion theory analysis is that of the interaction measured across a pure ionic liquid, of which there have been many over recent years\cite{gebbie2017,fung2023_il}; further examples are presented later in this manuscript. In these measurements, two decay modes are again observed: somewhat like the dilute electrolyte case, a monotonic decay dominates at long range, while an oscillatory decaying mode dominates at shorter range.  The oscillatory part typically has a wavelength close to the ion pair dimension, and so is interpreted as `ion layering' or, equivalently, an oscillatory mode arising predominantly from charge-charge correlations. The monotonic part of the interaction, which is enormously longer-ranged than can be explained from the Debye-H\"uckel picture alone\cite{smith2016_underscreening}, has been called `underscreening' and has been the subject of some debate\cite{Holm2020,coles2020,gaddam2019_ducker,kumar2022_absence,Hartel2023}. Within the framework of the dressed ion theory, we understand that a monotonically decaying mode can arise in an ionic liquid when the \textit{effective} charge is low, and recalling that this effective charge comes from the non-linear part of the screening cloud which has the size of the decay length. Physically, this could arise from strong ion correlations (i.e. strong nonlinearity), or associations, of the sort intuited by Pincus and Safran\cite{Safran_SM2023}. \\

A final example illustrating the insight obtained from this framework is the interaction across ionic liquids containing long-chain cations which are known from scattering experiments to be nanostructured in the bulk fluid. Neutron and X-ray scattering techniques reveal that several short-range structural modes contribute to the bulk structure of pure ionic liquids\cite{triolo2007,annapureddy2010,hettige2014,weththasinghage2020}{}: these arise from nearest neighbour interactions concerning the solvation shell around each ion; from charge ordering interactions, which determine the distribution of cations around cations and anions around anions at intermediate length scales; and from the segregation of the non-polar regions of molecular ions away from the charged regions, leading to nanostructured domains at a longer length scale. This third lengthscale in the bulk scattering typically emerges when non-polar chains on the ionic liquid ions are $>6$ carbon atoms long. Measurements of the surface forces across ionic liquids with increasing hydrocarbon chain length also reveal the appearance of this latter nonpolar nanostructural lengthscale: the measured oscillatory force across longer chain ionic liquids contains an oscillatory force with wavelength consistent with nonpolar ordering -- or `bilayers' as the repeat unit between the surfaces\cite{perkin2011_il,smith2013_bilayersa}. However, it is not clear why the nonpolar ordering dominates over charge ordering in the surface force measurement, and the interpretation of a sharp structural transition is unsatisfactory  for the implication of a phase instability which ought to give rise to an attractive interaction at some range -- which is not seen. Instead, we can interpret these measurements as follows. The long-chain ionic liquids support multiple modes in the bulk fluid, including modes arising from their charge- and geometric- asymmetry; some of these are revealed as peaks in the bulk scattering patterns. The surface force measurement picks up some \textit{but not all} of these modes, i.e. the pre-exponential factors in equation~\ref{eq:WI2D}, which are determined by the affinity of the surfaces for a particular mode, can be large for some modes and small or zero for others. That is, only some modes resonate with the SFB cavity.

From these examples, we see that the cavity between the mica sheets in the SFB creates a kind of electrostatic resonator, detecting some (but not all) of the modes (decay terms in equations~\ref{eq:hij} and ~\ref{eq:WI2D}) which define the decay of correlations and density in the bulk fluid. Different modes will dominate at different distances. Related to the requirement that a mode resonate with the mica cavity, in an SFB experiment the amplitudes of the measured modes can vary over a very wide range. We revisit this point later for its relevance to interpretation of the underscreening mode. 

In the remainder of the manuscript we present results from direct experimental measurements of the interaction force between atomically smooth mica plates across ionic liquids and their mixtures, measured using a surface force balance (SFB). As we shall see, the results are quite complex and certainly could not be interpreted in the simple physical manner that has been applied before for dilute electrolytes and some aspects of more concentrated electrolytes. Thus, their interpretation presents a useful challenge to the mode analysis outlined above. The interaction force directly gives the free energy of interaction and thus $W_{I,II}(D)$, and the elegant connections between (macroscopic) $W_{I,II}(D)$ and (microscopic) $h_{ij}(r)$, $g_{ij}(r)$ and $w_{ij}(r)$ proposed by the dressed ion theory provides a direct route to some decay modes and parameters which describe correlations in the bulk ionic fluids. We discuss and interpret our measured forces in this way and in doing so hope to better understand the relationship between structure and correlations within electrolytes and the interaction of macroscopic bodies across them. 

\section{Methods}

\begin{figure}[h!]
	\centering
	\includegraphics[scale=1]{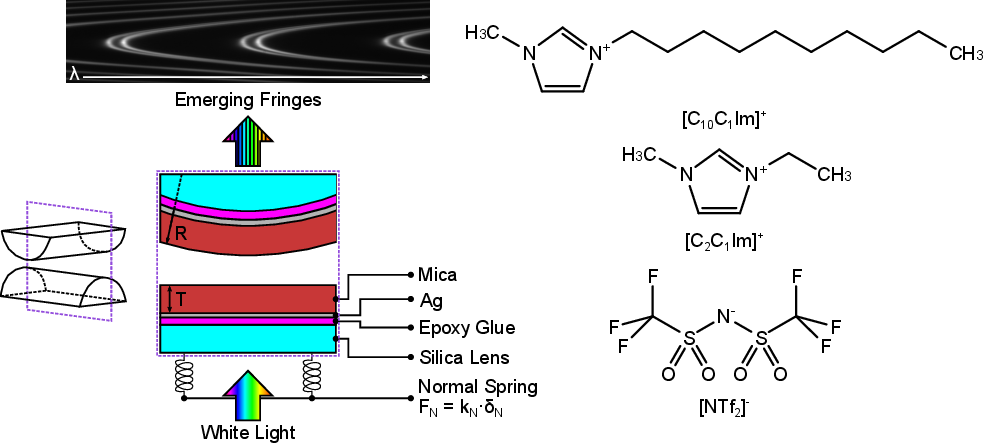}
	\caption{\textit{Left:} summary of the Surface Force Balance (SFB).
		Silver-backed mica sheets of identical thickness are glued, silver side down, onto two hemicylindrical lenses.
		The lenses are mounted vertically within the instrument in a crossed-cylinder configuration, with one lens mounted on a horizontal leaf spring with spring constant $k_N$.
		White light, normal to the surfaces, is shone through the resulting interferometric cavity.
		Interferences take place within the cavity, leading to the emerging Fringes of Equal Chromatic Order (FECO).
		Analysis of the FECO allows the surface separation $D$ and the radius of curvature of the lens $R$ to be found.
		\textit{Right:} The structures of the ionic liquids investigated in this study.
		The liquids share a common anion and differ by the length of the alkyl chain in the cation.}
	\label{fig:methods}
\end{figure}

We used the Surface Force Balance (SFB) to measure the interaction force as a function of separation distance between atomically smooth mica sheets across ionic liquids and ionic liquid mixtures. The apparatus and its operating principles has recently been described elsewhere\cite{groves2021_WiS,HaylerROPP2024}{}. The mica sheets are ruby muscovite (S\&J Trading Inc.), cleaved in a particle-free environment on both faces to give atomically smooth sheets of centimetric area and precisely uniform thickness across the whole area in the range $2<T_m<7$~\si{\micro\metre}. A silver layer of thickness $\sim$~45~nm is deposited onto the mica sheets, which are then cut to size and glued, silver side down, to a hemicylindrical glass lens with radius $R\sim$~0.01~m.  
Two lenses, prepared from the same mica sheet to ensure identical mica thickness on each lens, are then mounted vertically in the SFB instrument in a crossed-cylinder arrangement such that the mica surfaces face one another, as shown in figure~\ref{fig:methods}. The lower lens rests on a horizontal leaf spring of known spring constant $k_N$. Collimated white light is shone normal to the two lenses. The silver mirrors on the back of each mica surface form an interferometric cavity, with emerging light taking the form of a series of \textit{fringes of equal chromatic order} (FECO). Analysis of the FECO allows precise determination of the surface separation, with an accuracy of $\sim$~0.5~nm and a precision of $\sim$~0.1~nm.

The first stage of each experiment involves bringing the mica sheets into direct contact in dry air to measure (calibrate) the mica thickness, $T_m$. Subsequently, a fluid of interest is injected between the lenses, and the lenses are then approached or retracted in a linear fashion using a mechanical or piezoelectric drive. If interaction forces act between the surfaces, the spring upon which the lower surface is mounted deflects by a small amount $\delta_N$ which appears in the analysis of the FECO as a deviation from the applied linear motion. The interaction force  between crossed cylinders $F_N(D)$, can be calculated from Hooke's law.  $F_N(D)$ is related to the free energy of interaction per unit area at the same distance between parallel plates $G^{||}(D)$ by the Derjaguin approximation:
\begin{equation}
	G^{||}(D) = \frac{F_N(D)}{2 \pi R}
	\label{eq:derjaguin}
\end{equation}
where $R$ is the radius of curvature of the cylinders.This approximation holds when $R \gg D$, as is the case in all of the measurements we show here.

The SFB force runs reported here typically start from distances of $\sim$~200~nm, and the approach and retraction speeds are $<1$~nm~s$^{-1}$. Taking account of the viscosity of the ionic liquids, under these conditions we can interpret the forces measured as equilibrium forces (with no measurable velocity-dependent hydrodynamic contribution in this case)\cite{HaylerROPP2024}. Furthermore, if we (crudely) assume that the ions within the confined region between the surfaces have diffusion coefficients similar to in the bulk fluid, their characteristic time to travel over 100nm is $10^{-3}$s, compared to the approach time of the surfaces of $\sim10^2$s to cover the same distance. This implies that constituent ions explore all configurations during the approach, which is therefore quasi-static from an ergodic perspective as well as from the point of view of hydrodynamics. Combined with the macroscopic area investigated, we interpret the measured interaction free energy per unit area as equal to the potential of mean force per unit area between the two macroscopic plates across the fluid, i.e. $\frac{F_N(D)}{2 \pi R} = G^{||}(D) = W^{||}(D)$.

The ionic liquids studied are 1-ethyl-3-methylimidazolium bis(trifluoromethanesulfonyl)imide, \ce{[C2C1Im][NTf2]} (Iolitec, $> 99.5$~\%), and 1-decyl-3methylimidazolium bis(trifluoromethanesulfonyl)imide, \ce{[C10C1Im][NTf2]} (Iolitec, $> 98$~\%). Ionic liquids and mixtures were dried for approx. 24h before each experiment on a Schlenk line at $10^{-1}$mbar and +70$^\circ$C. This typically results in water content below 200ppm. The two ionic liquids have a common anion and a smilar cation headgroup, therefore varying the mole fraction in mixtures acts only to alter the fraction of shorter and longer alkyl chains in the ionic fluid.  The molecular structures of the ions is shown alongside the schematic of the SFB in figure~\ref{fig:methods}.
Experiments were carried out for mixtures of the two ionic liquids at \ce{[C2C1Im][NTf2]} mole fractions of 0.0, 0.1, 0.3, 0.5, 1.0. Each experiment involves a freshly cleaved pair of mica sheets. Within each experiment multiple experimental runs are made over several hours, including runs at different contact spots on the mica sheets. Fitted values in the following section arise from averages across multiple runs. For clarity in presentation of the results we show single example measurements in the figures within this manuscript, however all measurement runs, from which the averages were obtained, are available within the supporting data files [Here we will provide link to Oxford Research Archive Data deposit] 

\section{Results and Discussion}
\textbf{Pure Ionic Liquids.} 

\begin{figure}[h!]
	\centering
	\includegraphics[scale=1]{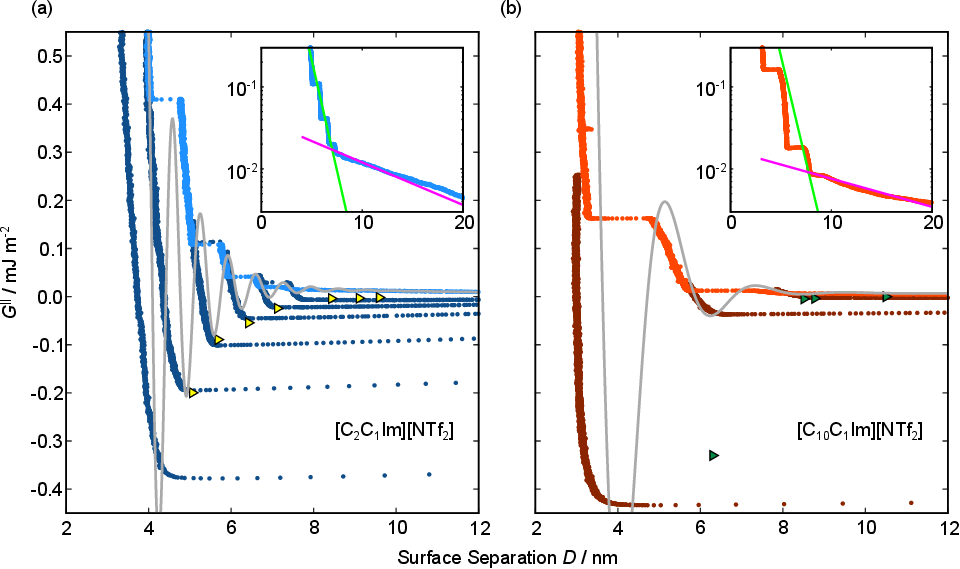}
	\caption{Interaction force, $F_N(D)$, between mica sheets at separation distance $D$ across pure ionic liquids; (a)  \ce{[C2C1Im][NTf2]} and (b) \ce{[C10C1Im][NTf2]}. The force is normalised by radius of curvature, $R$, of the crossed-cylinders and plotted as $G^{||}(D) = \frac{F_N(D)}{2 \pi R}$. Interactions measured on approach of SFB surfaces are shown in light colours, while those measured on retraction are shown in darker colours. Individual measurements of the minimum positions in each case are shown as triangles. Insets show the interactions on a a semi-log plot. The grey line shows a fit to the force profile as described in equation~\ref{eq:interaction_pure}, while the green and magenta lines shown in the inset highlight the exponential decay of the oscillatory and monotonic modes respectively. We note that small changes to the method of fitting the oscillatory force can lead to large variation in the amplitude of the oscillatory term. Here, we have used the energy minima to pin the fit, and errors mentioned in the text relate to this method. See SI for further details about the fitting procedure.}
	\label{fig:pureIL}
\end{figure}

Measurements of the interaction force as a function of separation distance between mica sheets across one short-chain ionic liquid, \ce{[C2C1Im][NTf2]} and one longer chain ionic liquid, \ce{[C10C1Im][NTf2]}, are presented in figure~\ref{fig:pureIL}(a) and (b), respectively, with the measured force plotted as $ \frac{F_N(D)}{2 \pi R} =G^{||}(D)$. Qualitatively, the results for the two pure ionic liquids show similar features: in each case, we observe an oscillatory region in the profile at surface separations between about 3 and 10~nm, defined by a series of repulsive walls and attractive wells measured during approach and retraction of the mica surfaces. The oscillations appear to decay exponentially as the surface separation is increased. At surface separations greater than about 10~nm the interaction is dominated by a longer-range monotonic component which also appears to decay exponentially as the surface separation is increased (see inset semi-log plot).

The observation of two distinct decay modes, one damped oscillatory and one plain exponential, is reminiscent of many previous measurements of interaction forces in pure ionic liquids\cite{horn1988_il,perkin2010_il,perkin2011_il,smith2013_SFBil,gebbie2013,gebbie2015,smith2016_underscreening,gebbie2017,fung2023_il}{} and can be interpreted as two dominating terms contributing to the overall potential of mean force as in equation~\ref{eq:WI2D}. We therefore parameterise the measured interaction as follows:
\begin{equation}
	G^{||}(D) = A~e^{-D/\lambda_o}~\text{cos}\bigg(\frac{2 \pi}{\xi}(D+\delta)\bigg) + B~e^{-D/\lambda_s}
	\label{eq:interaction_pure}
\end{equation}
where the first term in the right hand side of equation~\ref{eq:interaction_pure} describes the short range oscillating exponential decay mode, with magnitude $A$, exponential decay length $\lambda_o$, oscillatory wavelength $\xi$, and oscillatory offset $\delta$; and the second term describes the long range monotonic exponential decay mode, with magnitude $B$ and exponential decay length $\lambda_s$.
Note that the experimental decay lengths $\lambda_o$, $\lambda_s$ and wavelength $\xi$ mirror the theoretical decay lengths $1/\kappa_1$, $1/\kappa_R$ and wavelength $2\pi/\kappa_I$ from equation~\ref{eq:hij}; we maintain separate notation for the experimentally fitted values so as not to imply any exact equality to the ideal theoretical expressions. The fits to equation~\ref{eq:interaction_pure} are shown superimposed on the data in figure~\ref{fig:pureIL} and the fitted values for each ionic liquid are given in table~\ref{tab:fits_pure}. (See SI figure 1 for an example of the fitting procedure.)

\begin{table}[h]
	\centering
	\caption{Fitted parameters from equation~\ref{eq:interaction_pure} for the pure ionic liquids studied in this investigation.
		Errors are taken as 95~\% confidence intervals for $A$, $\lambda_o$ and $\xi$, as uncertainty in absolute $D=0$ position for $\delta$, and as standard deviations for $B$ and $\lambda_s$.}
	\label{tab:fits_pure}
	\begin{tabular}{|l||cc|}
		\hline
		Parameter & \multicolumn{1}{l|}{\ce{[C2C1Im] [NTf2]}} & \multicolumn{1}{l|}{\ce{[C10C1Im] [NTf2]}}  \\ \hline
		$A$ / mJ m$^{-2}$       &  100 $\pm$ 60       & 92 $\pm$ 69                
		\\
		$\lambda_o$ / nm        &  0.81 $\pm$ 0.10    & 0.84  $\pm$ 0.17           \\
		$\xi$ / nm              &  0.67 $\pm$ 0.03    & 2.20  $\pm$ 0.15           \\
		$\delta$ / nm           &  0.1 $\pm$  0.5     & 1.3 $\pm$ 0.5              \\
		$B$ / mJ m$^{-2}$       &  0.040 $\pm$ 0.011  & 0.017 $\pm$ 0.007         
		\\
		$\lambda_s$ / nm        &  8.33 $\pm$ 2.40    & 12.68  $\pm$ 3.43          \\
		\hline
	\end{tabular}
\end{table}

As discussed in the introduction, intuitive rationalisation of oscillatory forces across short-chain ionic liquids (as in fig.~\ref{fig:pureIL}(a)) in the past has been based on the close similarity of the oscillatory wavelength, $\lambda_o$, to the cation-cation or anion-anion correlation length in the bulk fluid. For \ce{[C2C1Im] [NTf2]}, the measured wavelength of 0.67~nm is close to the mean ion pair diameter of $\approx 0.75$~nm, (equivalent to the charge-charge correlation length). The oscillations are interpreted as arising from sequential squeeze-out of repeat-units consisting of layers of cations and anions. For the longer chain ionic liquids, as in fig.~\ref{fig:pureIL}(b), the wavelength of 2.20~nm is much longer than the mean ion pair diameter ($\approx 0.87$~nm) and matches more closely the nonpolar-nonpolar correlation length. This seems to imply sequential squeeze-out of ionic liquid bilayers comprised of cations arranged in tail-to-tail bilayer repeat units\cite{perkin2011_il,smith2013_bilayersa,Freitas2018}{}.  

Alternatively, the surface interactions in fig.~\ref{fig:pureIL}(a) and (b) could be described in terms of the dominating modes, i.e. those terms which resonate most strongly in the z-direction of the SFB cavity. For \ce{[C2C1Im] [NTf2]}, the dominating mode between mica surfaces at 3-10 nm separation is the charge-charge correlation mode, whereas for \ce{[C10C1Im] [NTf2]} the dominating mode in the SFB cavity is the nano-structuring mode - we observe the mode that corresponds to the organisation of the amphiphilic cation nonpolar tails. Other structural modes exist in the $h_{ij}(r)$, for example the charge-charge correlation and the density-density correlation of headgroups in \ce{[C10C1Im] [NTf2]}, but these do not resonate in the cavity -- i.e. their respective prefactors in eq.~\ref{eq:WI2D} are small -- and so are not picked up in the measurement.  
It is interesting to consider why, from a molecuar perspective, the nonpolar correlations dominate over the charge-charge correlations for the long chain ionic liquids. In a self-assembled bilayer structure the charge-charge correlation is in the plane of the bilayer while the nonpolar correlation is perpendicular to the plane. Therefore, in the cavity of the SFB, which detects modes resonating perpendicular to the mica surfaces, the fact that only one of these modes is picked up in the measurement indicates that the nanostructure is composed to some large extent of bilayer-like structures arranged parallel to the surfaces. 

The long range mode that is observed in both pure ionic liquids, as shown in the insets of figure~\ref{fig:pureIL}, is attributed to underscreening\cite{gebbie2013,Lee_FD2017}{}, as mentioned in the introduction. We return to discuss this later, in light of results for the mixtures.\\ 
~\\
\textbf{Ionic Liquid Mixtures}

\begin{figure}[h!]
	\centering
	\includegraphics[scale=1]{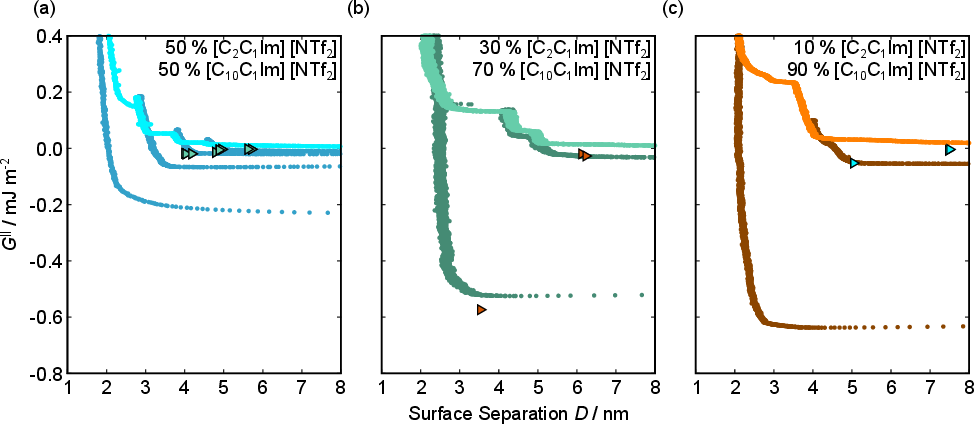}
	\caption{Interaction-distance profiles of ionic liquids mixtures.
		Shown are measurements for (a): 50~mol~\%, (b): 30~mol~\%, and (c): 10~mol~\% \ce{[C2C1Im][NTf2]} in \ce{[C10C1Im][NTf2]}.
		Interactions measured on approach of the SFB surfaces are shown in the lighter colour while retractions are shown in the darker colour.
		Individual measurements of minima positions are shown as triangles.
		}
	\label{fig:mixIL_nofit}
\end{figure}

Figure~\ref{fig:mixIL_nofit} shows the interaction-distance profiles collected in the ionic liquid mixtures.
In each mixture, we again observe two well defined regions: at surface separations of $<10$~nm, we see a series of attractive and repulsive interactions. At surface separations $>10$~nm the interaction is dominated by a longer range monotonic repulsive interaction, as will be shown later.

The short-range forces across these ionic liquid mixtures cannot be described by a single damped oscillating mode. In the case of 50~mol~\% \ce{[C2C1Im] [NTf2]} the oscillations appear to have an approximately uniform wavelength, but as the concentration of \ce{[C2C1Im] [NTf2]} is reduced to 30 and then 10~mol~\%, the form of the measured interaction profiles become much more complex. In each case, the interaction profile shows non-monotonic trends in the positions of maxima and minima. Distances between maxima and minima are also non-uniform, and may be smaller than 0.5~nm or larger than 1.5~nm, within the same profile. We emphasise that, \textit{despite their irregular form, these measurements were found to be highly reproducible over $\sim 12$~hours of measurement time}; the features of of 30 and 10~mol~\% \ce{[C2C1Im] [NTf2]} are shown in more detail in the supporting information.

\begin{figure}[h!]
	\centering
	\includegraphics[scale=1]{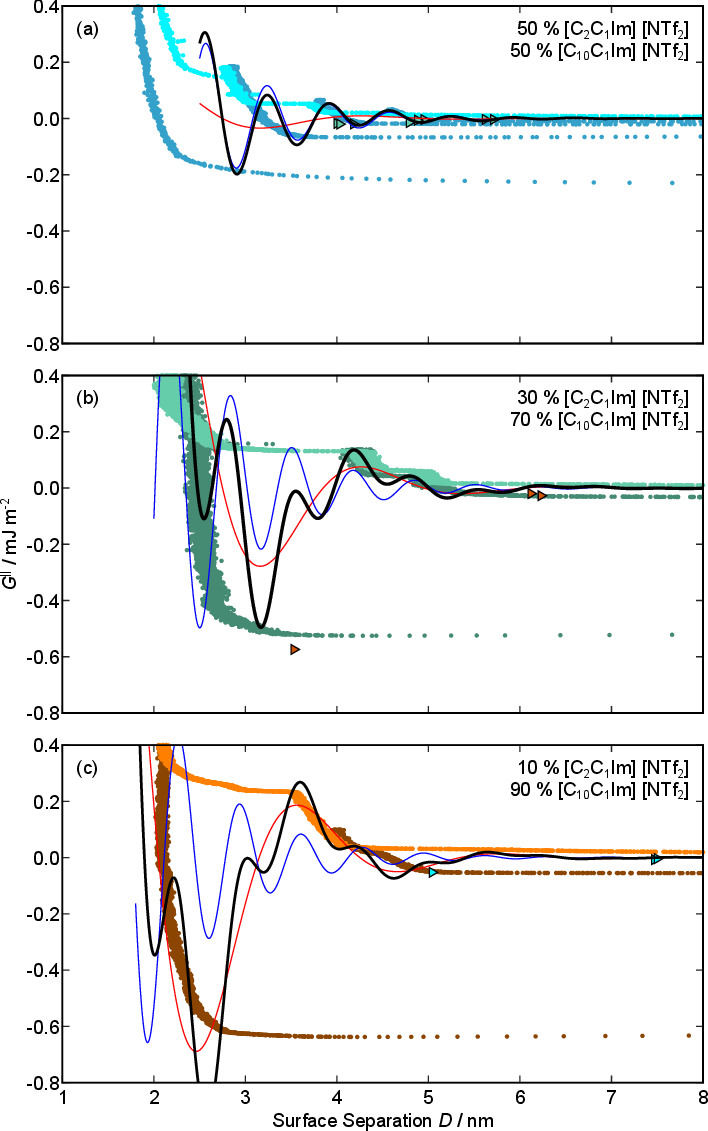}
	\caption{Interaction-distance profiles of ionic liquids mixtures with fits as described in equation~\ref{eq:mix_fit} overlaid.
	The total fit is shown in black, with the contribution to the total fit from the charge ordering mode of \ce{[C2C1Im] [NTf2]} shown in blue and the contribution from the bilayer mode of \ce{[C10C1Im] [NTf2]} shown in red.
	}
	\label{fig:mixIL_fit}
\end{figure}

Clearly, we cannot rationalise the form of these irregular repulsions and attractions using intuitive arguments about ion layers and bilayers. However, we can attempt a model for the measurement based on the expected decay modes present in the bulk liquid mixture.  To reiterate, the bulk fluid will be characterised by correlation functions, $h_{ij}(r)$, containing many terms, of the monotonic Yukawa or oscillatory Yukawa types only, as in equation~\ref{eq:hij}. Some of these modes will be similar to the modes present in the pure ionic liquids (e.g. cation-anion density correlations), while some will be particular to the mixture (e.g. cation--cation correlations between the two types of cation). Of these, some (but not all) modes will appear in the measurement of surface forces. A reasonable starting point for fitting the data is to assume that each of the oscillatory modes present in SFB measurements for pure ionic liquids may contribute to interactions in the mixture. We take only a single monotonic mode, since the data show no hint of two plain exponential decays; the monotonic underscreening mode is discussed later. Thus we write as a trial function:

\begin{equation}
	G^{||}(D) = M~e^{-D/\lambda_o'}~\text{cos}\Bigg(\frac{2 \pi}{\xi'}(D + m)\Bigg) +  N~e^{-D/\lambda_o''}~\text{cos}\Bigg(\frac{2 \pi}{\xi''}(D + n)\Bigg)+Be^{-D/\lambda_s}
	\label{eq:mix_fit}
\end{equation}
where the superscript $'$ refers to fixed parameters taking the values from pure \ce{[C2C1Im][NTf2]} and $''$ to fixed parameters calculated for \ce{[C10C1Im][NTf2]}, as given in table~\ref{tab:fits_pure}. The first term in equation~\ref{eq:mix_fit} represents the contribution to the interaction from the charge ordering mode of \ce{[C2C1Im][NTf2]} and the second term the contribution from the bilayer mode of \ce{[C10C1Im][NTf2]}. $M$ and $N$ are fitting parameters describing the magnitude of each oscillatory term, and therefore the `weight' of the contribution of each pure liquid to the overall interaction-distance profile. $m$ and $n$ are offsets for each oscillatory term. The parameters for the third term on the right, the monontonically decaying mode (underscreening), are discussed later. The best fit values for $M$, $N$, $m$, and $n$ are determined manually and are given in table~\ref{tab:mixfits} and shown superimposed on the experimental measurements in figure~\ref{fig:mixIL_fit}.

\begin{table}[h]
	\centering
	\caption{Fitting parameters from equation~\ref{eq:mix_fit} for forces measured in ionic liquid mixtures.
			Values for the pure liquids from equation~\ref{eq:interaction_pure} are shown for comparison (\textit{i.e.} $A$ in table~\ref{tab:fits_pure} is $M$ for 100\% \ce{[C2C1Im] [NTf2]} and $N$ for 0\% \ce{[C2C1Im] [NTf2]}. Note that $\lambda_o'$, $\lambda_o''$, $\xi'$, and  $\xi''$ are not fitting parameters but are fixed at the values determined for the pure ionic liquids as in in table~\ref{tab:fits_pure}}
	\label{tab:mixfits}
	\begin{tabular}{|l||ccccc|}
		\hline
		Mole \%& \multicolumn{1}{l|}{$M$}& \multicolumn{1}{l|}{$m$} & \multicolumn{1}{l|}{$N$} &  \multicolumn{1}{l|}{$n$} & \multicolumn{1}{l|}{$M/N$} \\
		\ce{[C2C1Im] [NTf2]} & \multicolumn{1}{l|}{/ mJ m$^{-2}$}  & \multicolumn{1}{l|}{/ nm} & \multicolumn{1}{l|}{/ mJ m$^{-2}$}  & \multicolumn{1}{l|}{$/ nm$} & \\ \hline
		100  & 100   & 0.1    &   0  &  0   & N/A   \\
		50   &  6.4  & 0.09   &  1.6 &  0   & 4     \\
		30   &  11   & 0.5    &  13  &  0   & 0.85  \\
		10   &  7.2  & 0.4    &  14  & 0.7  & 0.51  \\
		0    &   0   & 0      &  92  & 1.3  & N/A   \\
		\hline
	\end{tabular}
\end{table}

Remarkably for such a simple model, the fits are able to capture much of the complex behaviour of the recorded interaction profiles. For 50~mol~\% \ce{[C2C1Im][NTf2]}, the major contribution comes from the charge-charge (shorter wavelength) term, although the amplitude is substantially less than for pure \ce{[C2C1Im][NTf2]}. The non-monotonic progression of maxima and minima seen at 30 and 10~mol~\% \ce{[C2C1Im][NTf2]} are reproduced when both the charge-charge mode and the nonpolar ordering (bilayer) modes have similar magnitudes.\\
One interesting feature of these fits is that the magnitudes of the fitted forces in the mixed liquids at all concentrations investigated are an order of magnitude lower than those calculated for the pure liquids. This is similar to the observation made in a mixture of an ionic liquid with a polar solvent\cite{smith2017_switching}{}, in which the magnitude of the observed oscillations also fell at intermediate concentrations. That is to say, the structural modes can interfere destructively.\\
Unfortunately, the absolute values of the magnitudes of forces, even in the pure liquids, are highly variable between measurements; small perturbations in twist angle of the two mica sheets, in concentration and in surface chemistry can lead to large changes in the values of $A$ in equation~\ref{eq:interaction_pure}. For this reason it is difficult to extract information from the absolute fitted values $M$ and $N$, however it can be instructive to look at the ratio $M/N$ for a single experiment, also given in table~\ref{tab:mixfits}, where it is clear that $M/N$ decreases as the mole fraction of \ce{[C2C1Im][NTf2]} is decreased. This makes intuitive sense - the contribution to the overall interaction potential from the modes of the liquid that is in the highest concentration dominate. This does not seem to track linearly with concentration, and at equimolar concentrations it is the charge-charge mode of \ce{[C2C1Im][NTf2]} that dominates the interaction. This perhaps suggests that the affinity of this mode for the surfaces is greater than that of the bilayer mode of \ce{[C10C1Im][NTf2]}, which may be interpreted in the molecular picture as a greater ability of the long chain ionic liquid to dissolve in the short chain liquid than vice versa. Perhaps relatedly, scattering experiments in similar liquid mixtures also revealed multiple modes including the nonpolar ordering mode at high concentrations of the longer chain ionic liquid\cite{cabry2018,weber2019,cabry2022_slattery}{}.

\begin{figure}[h!]
	\centering
	\includegraphics[scale=1]{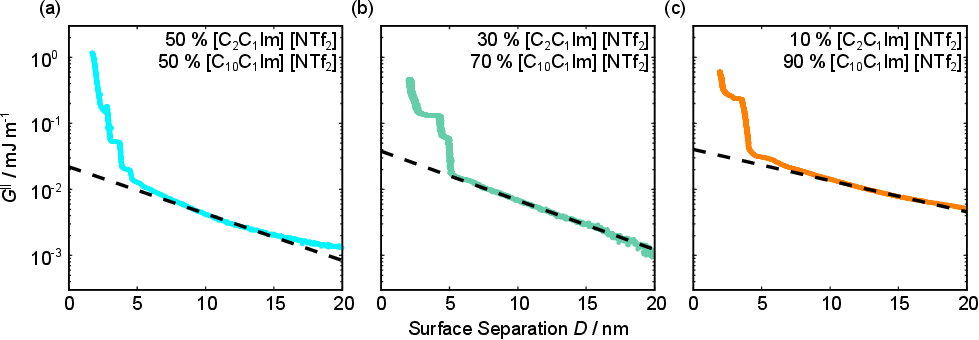}
	\caption{Interaction-distance profiles of ionic liquids mixtures with long range exponential fits, plotted on a logarithmic energy scale.
	}
	\label{fig:mixIL_longrange}
\end{figure}

We now turn to consider the long range plain exponenetial decay measured in the mixtures, represented by the final, monotonic, mode in equation~\ref{eq:mix_fit}. Measured forces in the ionic liquid mixtures are shown on a semi-log scale and extending to greater $D$-values in figure~\ref{fig:mixIL_longrange} in order to facilitate inspection of the longest decaying mode. For each of the mole ratios studied, as for the pure ionic liquids, there is a clear monotonically decaying force beyond the oscillatory force extending beyond 10~nm. Fitting these modes to the monotonic term in equation~\ref{eq:mix_fit} gives the values as listed in table~\ref{tab:mixfits_u}, where values for the pure ionic liquids are also provided for comparison.

\begin{table}[h]
	\centering
	\caption{Fitting parameters for the final, monotonic term in equation~\ref{eq:mix_fit} for forces measured in ionic liquid mixtures and for pure ionic liquids (from equation~\ref{eq:interaction_pure}) for comparison.
		Errors are taken as standard deviations from fits on multiple runs.}
	\label{tab:mixfits_u}
	\begin{tabular}{|l||cc|}
		\hline
		Mole \% \ce{[C2C1Im] [NTf2]}& \multicolumn{1}{l|}{$B$ / mJ m$^{-2}$}& \multicolumn{1}{l|}{$\lambda_s$/ nm} \\ \hline
		100  & 0.040 $\pm$ 0.011 & 8.33 $\pm$ 2.40    \\
		50   & 0.025 $\pm$ 0.007 & 6.37 $\pm$ 0.77    \\
		30   & 0.036 $\pm$ 0.007 & 8.99 $\pm$ 2.07    \\
		10   & 0.054 $\pm$ 0.020 & 9.87 $\pm$ 2.32    \\
		0    & 0.017 $\pm$ 0.007 & 12.68 $\pm$ 2.32   \\
		\hline
	\end{tabular}
\end{table}

The slowly-decaying monotonic mode in ionic liquid mixtures is likely of similar origin to that reported many times for similar measurements in pure ionic liquids and discussed above; this mode has been called \textit{underscreening} or \textit{anomalous underscreening}\cite{Hartel2023} in recognition that the charged surfaces appear less electrostically screened than predicted by mean field theories. The underscreening decay lengths range from about 6~nm to 12~nm, although there is scatter in both $B$ and $\lambda_s$ (see table). 

One notable observation is that, in the mixtures, only a single underscreening decay mode is observed. This contrasts with the oscillatory region of the interactions in mixtures where two oscillatory modes are superimposed. This is perhaps related to the longer-range effects which give rise to underscreening; the ion correlations which drive non-linear contributions to the correlation functions are determined by charge density distribution, but are less sensitive to the molecular details of the ions. 

Another notable feature of the underscreening mode seen here for mixtures, but also for pure ionic liquids, is the very small amplitude (pre-exponential factor) of the term compared to the amplitude of oscillatory modes present in the same system. For example, comparing the numerical values of $B$ (underscreening amplitudes) in table~\ref{tab:mixfits_u} with the amplitudes of the oscillatory terms $A$, $M$ and $N$ in table~\ref{tab:mixfits} we see that -- despite large scatter -- the underscreening is typically a factor of $10^3$ weaker. To detect in a single experiment or simulation both the short-range oscillatory mode, which has a short wavelength and very large amplitude, and a long range monotonic mode, with long wavelength (by factor of $>10$) and much smaller amplitude (by factor of  $10^3$--$10^4$) requires sufficiently high resolution over a particularly large dynamic range. This may explain why, in some cases, the underscreening mode has not been picked up in some experiments or simulations. 
\section{Conclusions}
We have presented direct measurements of the force-distance profiles between mica sheets across the ionic liquids \ce{[C2C1Im][NTf2]} and \ce{[C10C1Im][NTf2]}, and their mixtures, using a surface force balance. In the pure liquids, two modes were observed: a long range monotonic mode, characteristic of the underscreening interaction seen in high concentration electrolytes, and a short range damped oscillatory mode. For \ce{[C2C1Im][NTf2]} the oscillatory mode had the wavelength close to the charge-charge correlation length in the bulk fluids, while in \ce{[C10C1Im][NTf2]} the oscillatons corresponded to a bilayer self-assembly mode. In the mixed liquids, again at large surface separations a monotonic mode characteristic of underscreening was observed. However, at small separations the series of potential maxima and minima could no longer be fitted to a single decaying oscillation. Instead, we considered a linear superposition of the short-wavelength charge ordering mode from \ce{[C2C1Im][NTf2]} and the long-wavelength bilayer mode from \ce{[C10C1Im][NTf2]}; this simple model was able to reasonably capture the behaviour of the observed interaction-distance profiles. We introduced our work with a brief overview of how the dressed ion theory of Roland Kjellander, which provides an exact treatment of ion correlations, can be useful in describing the various contributions to the interaction free energy measured in this sort of experiment. Key messages from the theoretical side, such as the multiplicity of decay modes in all electrolytes and the necessity that these modes follow either exponentially decaying or oscillatory exponential forms, were pointed out and applied in describing the experiments. The same modes characterise the bulk correlation function, the decay of potential, and the potential of mean force between particles and also between macroscopic surfaces such as in the SFB. However the SFB cavity may only resonate with some of the modes; the manner in which the interfaces pin the waves and the possibility for wave superposition indicates a pleasing sort of wave mechanics mediated by the electrolyte solutions. \\

\begin{acknowledgement}
The authors would like to thank the European Research Council for financial support under grant 101001346, ELECTROLYTE.
TG is also grateful to the Worcester College Taylor Scholarship. Amaar Sardharwalla is thanked for assistance with viscosity measurements. Roland Kjellander, Sam Safran and Fyl Pincus are thanked for many delightful and insightful discussions. 
\end{acknowledgement}

\bibliography{bibliography}

\end{document}